\documentclass[english,aps,pra,showpacs,twocolumn]{revtex4}
\usepackage{graphicx}
\usepackage{babel}
 

\begin{document}

\title{Tunneling in Fractional Quantum Hall line junctions}

\author{M.~Aranzana}

\author{N.~Regnault}

\author{Th.~Jolic{\oe}ur}

\affiliation{Laboratoire Pierre Aigrain, ENS, D\'epartement de
Physique, 24 rue Lhomond, 75005 Paris, France}


\begin{abstract}
We study the tunneling current between two counterpropagating edge
modes described by chiral Luttinger liquids when the tunneling
takes place along an extended region.
We compute this current perturbatively by using a tunnel Hamiltonian.
Our results apply to the
case of a pair of different two-dimensional electron gases in the
fractional quantum Hall regime separated by a barrier, e. g.
electron tunneling. We also discuss the case of strong interactions
between the edges, leading to nonuniversal exponents even in the case
of integer quantum Hall edges.
 In addition to the expected nonlinearities due to
the Luttinger properties of the edges, there are additional
interference patterns due to the finite length of the barrier.
\end{abstract}

\pacs{73.43.-f, 71.35.Lk, 71.23.An}

\maketitle



\section{Introduction}
Edge states in the fractional quantum Hall effect (FQHE) are very
interesting examples of one-dimensional strongly interacting
quantum systems. The right and left moving edge modes of a quantum
Hall bar are spatially separated by a macroscopic length and this
leads to exponentially small backscattering. The main reason for
localization is thus suppressed and the edge modes are a nearly
ideal ballistic system. Many experiments have been devoted to the
study of their unique characteristics~\cite{Chang03}. An
interesting geometry is a constriction of the electron gas, the
so-called quantum point contact. By means of an electrostatic
potential created by a gate, the two edges of a bar are brought in
close proximity, allowing tunneling phenomena to take place at a
single point of the fluid. Recently, progress in the technique of
cleaved-edge overgrowth~\cite{Kang00,Grayson} has led to the fabrication of samples in
which the tunneling now occurs along a barrier of mesoscopic
extent between two spatially separated two-dimensional
electron gases. Kang et
al.~\cite{Kang00,Kang03} have performed detailed studies of the
conductance of these new structures. Their samples consist of
two-dimensional electron gases (2DEGs)  separated by
an atomically precise barrier of length 100 $\mu$m and of width 8.8 nm.
They have studied
the conductance of this structure as a function of the applied
magnetic field and the voltage bias between the two gases. Many
theoretical works have  tried to explain their
results~\cite{Takagaki00,Mitra01,Lee01,Kollar02,Nonoyama02,
Nonoyama02-2,Kim03,Kim03-2,Kim04,Yang04}.


In this paper, we give the results of a perturbative calculation
of  tunneling between two edge modes that are counterpropagating.
Each of these edge modes are described by a chiral Luttinger
liquid and we focus on the situation where they are characterized
by the same anomalous exponent $g$. We consider the situation
where tunneling takes place along an extended region with constant
amplitude.   The anomalous
exponent $g$ of the chiral Luttinger liquid is governed by the
bulk FQHE fluid(s) and enter in the expression of the correlation
function of the particle that tunnel. The extended tunneling
geometry is potentially rich of new interference phenomena not
found in single point-contact devices. Notably, T.~L.~Ho pointed
out the existence of oscillating currents without AC drive in the
case of integer edge modes\cite{Ho94}. We find that in the FQHE
regime there is a nontrivial interplay between the well-known
nonlinearities of the Luttinger liquid and the interference
pattern of the tunnel barrier. We also discuss the case of strong
interactions between the left and right-moving modes (but still weak
tunneling) which is amenable to the same theoretical treatment.
The exponent $g$ then takes nonuniversal values dictated by the details
of the interaction potential, even for integer quantum Hall edges.
  The formulas we obtain
are generalizations of previous results already available in the
literature~\cite{Wen91,Governale00}. This geometry may also give
rise in the presence of disorder to a delocalization
transition\cite{Renn95,Kane97}.
The line junction geometry may also be relevant to a recent set of experiments
using samples with an extended constriction~\cite{Roddaro}.
A detailed modelling of such line junctions has been performed
by Papa and McDonald~\cite{Papa04,Papa05}.

In section II, we introduce a model of edge states in the quantum
Hall regime with an extended tunnel barrier. In section III, we present
the perturbative treatment of the weak tunneling regime. Section IV gives the
results of our study. Finally section V contains our conclusions.

\section{The geometry and kinematics of tunneling}

\begin{figure}[!htbp]
\begin{center}\includegraphics[width=3.25in,
 keepaspectratio]{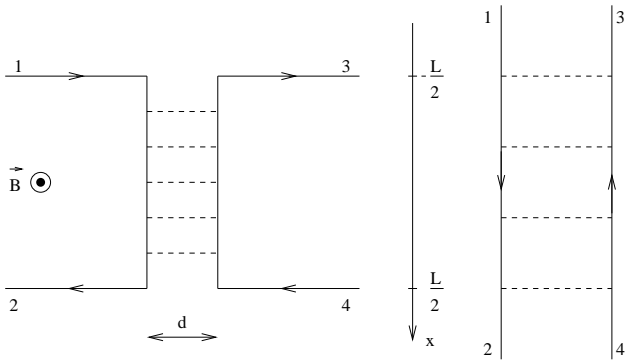}
\end{center}
\caption{Two laterally separated 2DEGs
under a magnetic field. A tunnel barrier causes mixing of the
counterpropagating edge modes. Realistic configuration (a),
topologically equivalent geometry (b)} \label{sample}
\end{figure}

The geometry we consider is illustrated in
fig.~\ref{sample}. It is that of an extended barrier whose height
remains constant along some spatial extent we call $L$. The
barrier separates spatially two 2DEGs. This is the
geometry of the samples studied by Kang et al.~\cite{Kang00}. In
the quantum Hall regime there are edge modes of conduction that
are counterpropagating. With the atomically precise barrier of
ref.~\cite{Kang00}, there is tunneling only in the
\textit{integer} quantum Hall regime when the applied bias between
the two 2DEGs is small. This geometry may also be realized by
electrostatic gates, in which case the energetics may be different
and extended tunneling may eventually be realized also in the
fractional quantum Hall regime.

We first discuss the kinematics of tunneling by reasoning in the
case of bulk filling factor $\nu =1$. This
situation was first investigated by T.~L.~Ho~\cite{Ho94}. The
Landau levels are degenerate in the bulk and their energies raise
when they approach a barrier. This scheme applies to both sides of
the barrier. Ultimately the Landau levels of both sides will cross
in the forbidden region inside the barrier and tunnel effect will
lead to the opening of single-particle gaps. A simple effective
theory clearly displays these phenomena. We first introduce the
left and right moving chiral fermions $\psi_{R}$ and $\psi_{L}$
that are the relevant modes when the bulk filling factor is
between $\nu =1$ and $\nu =2$. They are the counter-propagating
edge states~\cite{Wen91}. Their kinetic energy purely due to the
confining potential is then~:
\begin{equation}\label{Free}
\mathcal{H}_{kin}=-i v \int
dx(\psi^{\dagger}_{R}\partial_{x}\psi_{R}-\psi^{\dagger}_{L}\partial_{x}\psi_{L}),
\end{equation}
where $v$ is the drift velocity along the barrier. In momentum
space the left modes have energy $\epsilon_L (k)=-vk$ and the
right modes have $\epsilon_R (k)=vk$. These dispersion relations
cross at $k=0$ for zero energy. It is at this point that the
tunnel effect is strongest. The tunneling through
the extended barrier can be described by a Hamiltonian mixing
these modes~:
\begin{equation}\label{tunnel}
\mathcal{H}_{tunnel}=\mathcal{T} \int_{-L/2}^{+L/2} dx\, \,
(\psi^{\dagger}_{R}(x)\psi_{L}(x)+\psi^{\dagger}_{L}(x)\psi_{R}(x)),
\end{equation}
where $\mathcal{T}$ is the tunnel amplitude, constant along the barrier
by hypothesis.
This is the tunneling Hamiltonian we treat in this paper. The
coordinate $x$ is defined as in fig.~\ref{sample} so that the
modes propagate for all values of $x$ and tunneling is restricted
to the range ($-L/2, +L/2$).

This Hamiltonian can be seen a mass term of the fermion field for
an infinite-length barrier. This mass leads to a gap in the free
fermion excitation spectrum and this gap implies the existence of
a range of energies for which there are no propagating solutions
but only evanescent waves along the barrier. This phenomenon
repeats in a real sample also at the crossings of the other
higher-lying Landau levels as discussed by Kang et al..

The problem defined by $\mathcal{H}_{kin}+\mathcal{H}_{tunnel}$ is
a free fermion theory and has been studied exactly~\cite{Ho94}.
The scattering properties of the barrier can be deduced by a
straightforward Landauer-type calculation. However this simplicity
does not extend into the realm of the FQHE~: here the electron
operators are described by a chiral Luttinger liquid ($\chi$LL)
theory and no longer by the free theory Eq.(\ref{Free}). We study
the case with no interactions through the barrier so the $\chi$LLs
are decoupled and all the properties of their correlation
functions are known. We still use the tunneling Hamiltonian
Eq.(\ref{tunnel}) and treat it in perturbation to obtain the
current - voltage $I-V$
characteristic of this system. We consider the case of a single
edge appropriate to a $\nu=1/m$, $m$ odd, FQHE fluid. A single
chiral boson is then enough to describe the $\chi$LL theory. The
bulk fluid determines the value of the anomalous exponent $g$
characterizing the correlations at the edge. When the tunneling is
between two distinct gases then the operators $\psi_{R}$ and
$\psi_{L}$ should be taken as \textit{electron} operators.
Contrary to the case of
the quantum point contact geometry, the bosonized version of the
theory cannot be treated exactly since the tunneling term is no
longer a boundary operator. We are thus left only with the
perturbative calculation.

In real samples, there may be also noticeable Coulomb
interactions along and across the barrier which can be modeled by
the following kind of Hamiltonian~:
\begin{eqnarray}\label{Coulomb}
\nonumber
H_{int}&=&\int dx dy\,\,
V_{1}(x-y)\, [{\rho_{R}(x)\rho_{R}(y)+\rho_{L}(x)\rho_{L}(y)}]\\&+&2\int
dx dy \,\, V_{2}(x-y)\, \rho_{R}(x)\rho_{L}(y),
\end{eqnarray}
where $\rho_{R}(x) = :\psi^{\dagger}_{R}(x)\psi_{R}(x): $ is the
right-moving edge electron density operator  describing density
fluctuation at point $x$ ($\rho_{L}(x)$ is similarly defined).
This may be potentially important in the structure of Kang et al
where the width of the barrier is smaller than the magnetic
length. In this case, Wen~\cite{Wen91} has shown that the
eigenmodes are a mixture of the uncoupled left and right moving
edge modes. He showed also that the study of the tunnel effect may
be performed along the standard lines with simply a redefinition
of the Luttinger parameter $g$ when the momentum dependence of the
interactions may be neglected. So the results of our perturbative
tunneling calculation also apply to interacting edges if we use an
appropriate value of $g$. For example, Mitra and
Girvin~\cite{Mitra01} have estimated the value of $g$ to be
$\approx 0.6 - 0.7$ in the structures studied in
refs.(\cite{Kang00,Kang03}). This value is an example of what can be expected
in the integer quantum Hall regime when interactions destroy the perfect quantization of
$g$ by the bulk physics. Strictly speaking, this analysis applies when
the interedge is translation invariant along the barrier, like in Eq.(\ref{Coulomb})
and the tunnel

We note that such perturbative calculations have been done in
various tunneling situations and are a valuable tool to perform
spectroscopy of reduced dimensionality
samples\cite{Governale00,Boese01,Zulicke02,Zulicke02-2,Zulicke04}
in the quantum Hall regime.

In a given sample under a magnetic field, the edge states will be
populated up to a definite Fermi level. We will consider the
symmetric situation where the Fermi level is the same on both
sides of the sample. Application of a bias voltage induces a
difference between these Fermi levels. The external magnetic field
may be tuned so that the Fermi points meet at $k=0$. This is the
case first observed by Kang et al where there is a zero-bias peak
in the conductance. The peculiar dispersion of the edge modes
leads to special kinematic constraints in the case of an extended
barrier. Indeed, when the length of the barrier goes to infinity,
momentum along the barrier is a conserved quantity during
tunneling (contrary to the case of point tunneling). Also energy
conservation in the tunnel effect means that only states at the
intersection of the dispersion relations of the left and right
moving modes will tunnel. For a nonzero Fermi wavevector (defined
with respect to the "vacuum" situation
Eqs.(\ref{Free},\ref{tunnel})) this implies that one needs a finite
bias $eV$ to get a tunneling current. The voltage threshold will
go to zero if the filling factor is tuned to the degeneracy point
$k_f =0$.

\section{Perturbative treatment}

We now describe the application of the standard tunneling
formalism to the line junction. Tunneling occurs from a
non-equilibrium situation in which the chemical potential of the
left edge $\mu_{L}$ is different from $\mu_{R}$ of the right side
because of the applied voltage $eV=\mu_{L}-\mu_{R}$. The current
is expressed as the rate of change of the number of particles
$\widehat{N}_{L}=\int dx(\psi^{\dagger}_{L}(x)\psi_{L}(x))$, for
example, on the left-hand side of the sample. Thus, the average
tunneling current is given by:
\begin{equation}\label{Idef}
I(t)=-e\langle\dot{N}_{L}(t)\rangle ,
\end{equation}
where $e$ is the electron charge.
We set $\hbar=1$ and $k_B=1$ everywhere.

The total Hamiltonian is of the form~:
\begin{equation}\label{totHam}
\mathcal{H}=\mathcal{H}_{L}+\mathcal{H}_{R}+\mathcal{H}_{tunnel},
\end{equation}
where  $\mathcal{H}_{L}$ and $\mathcal{H}_{R}$ are Hamiltonians
for the chiral Luttinger liquids on each side of the junction. We
treat $\mathcal{H}_{tunnel}$ as a perturbation and keep only the
leading term. Since we limit ourself to the tunneling at low
voltage, low temperature and small tunneling amplitude
$\mathcal{T}$, $\mathcal{H}_{tunnel}$ is the only term which does
not commute with $\widehat{N}_{L}$ as $\mathcal{H}_{L}$ and
$\mathcal{H}_{R}$ separately conserve $\widehat{N}_{L}$ and
$\widehat{N}_{R}$. Standard first order
expansion in the interaction representation
yields~\cite{MahanBook}~:
\begin{equation}
I(t)={\displaystyle i e}\int_{-\infty}^{t}
dt'\langle[\dot{N}_{L}(t),\mathcal{H}_{tunnel}(t')]\rangle .
\end{equation}

Following \cite{MahanBook}, this procedure leads to the following
formula for the tunneling current~:
\begin{eqnarray}\label{IV}
&&I={\displaystyle e \mathcal{T}^{2}}
\int \frac{\displaystyle dk dk'}{\displaystyle 4\pi^{2}}
\frac{\displaystyle \sin^{2}[(k-k')L/2]}{\displaystyle
(k-k')^{2}}\times \\\nonumber
&&\times
\int \frac{\displaystyle d\epsilon}{\displaystyle
2\pi}A_{R}(k,{\epsilon}
)A_{L}(k',{\epsilon-eV}
)\{n_{f}(\epsilon-eV)-n_{f}(\epsilon)\}.
\end{eqnarray}
In this equation $n_{f}(\epsilon)=1/(1+\exp(\beta \epsilon))$ is
the Fermi factor at temperature $T$ ($\beta=1/T$) and
$A_{R}(k,\omega)$ (resp. $A_{L}(k,\omega)$) is the chiral spectral
function for  the right  (resp. left) moving chiral Luttinger
liquid.

The spectral function may be obtained from the imaginary part of
the retarded Green's function~:
\begin{eqnarray}
A_{R,L}(k,\omega)&=&-2\,\,\mathbf{Im}\int\int\frac{\displaystyle dt
dx}{\displaystyle 4\pi^{2}} \,\, e^{i\omega t- ikx}\\[1mm]
\nonumber
&&\times \left[-i
\theta(t)\langle \{\psi_{R,L}(x,t),\psi^{\dagger}_{R,L}(0,0)
\}\rangle_{\beta}\right],
\end{eqnarray}
where we take the thermal average $\langle\rangle_{\beta}$ in a
grand canonical ensemble including chemical potentials
$\mu_{L,R}$. We first evaluate the Fourier transform in space and
time of the forward Green function $G^{>}_{R,L}$ of right-movers
and left-movers; then the spectral densities are obtained by use
of the following identity which holds for fermions at finite
temperature~:
\begin{equation}
G^{>}_{R,L}(k,\omega)=(1-n_{f}(\omega))A_{R,L}(k,\omega).
\end{equation}
The Green functions  for a chiral Luttinger liquid are given by~:
\begin{eqnarray}
\nonumber
G^{>}_{R,L}(k,\omega)= &&\alpha^{g-1}\int\int\frac{\displaystyle dt
dx}{\displaystyle 4\pi^{2}}\,\, e^{i\omega t- ikx}e^{\pm ik_{f}x}\\[5mm]
&&\times\frac{\displaystyle (\pi T/v)^{g}}{ \displaystyle [i\, \sinh(\pi T(
t \mp x/v))]^{g}},
\end{eqnarray}
where $g$ is the Luttinger liquid parameter~\cite{Wen91}, $k_f$ is
the fermi wavevector and $\alpha$ is a microscopic cut-off. In the
case of \textit{electron} tunneling, the anomalous exponent is
$g=1/\nu =m$ when the two FQHE fluids have filling $\nu =1/m$.
This will describe the situation where a barrier separates two
distinct electron gases. Here we consider only the simplest
situation with the same filling factor on both sides of the
barrier. If we consider tunneling with interactions through the
barrier, then $g$ is non-quantized, e. g. in an integer Hall
system it may be slightly smaller than one (due to repulsive
interactions).

The Fourier transforms of the spectral functions can be
computed~\cite{Gradshteyn}~:
\begin{eqnarray}\nonumber
G^{>}_{R,L}(k,\omega)= &&(\frac{\displaystyle 2\pi
\alpha}{\displaystyle v\beta})^{g-1} e^{ \beta \omega
/2}\, \delta (\omega + v( k_{f} \mp k))\\[5mm]
&&\times B( \frac{g}{2} + i \frac{\beta
\omega}{2\pi},\frac{g}{2} - i \frac{\beta \omega}{2\pi} ).
\end{eqnarray}
This leads to the following closed formula~:
\begin{eqnarray}\label{spectral}
\nonumber
A_{R,L}(k,\omega)= &&2(\frac{\displaystyle 2\pi
\alpha}{\displaystyle v\beta})^{g-1} \cosh
({\frac{1}{2}\beta \omega})\, \delta (\omega + v( k_{f} \mp k))\\
&&\times B( \frac{g}{2} + i \frac{\beta
\omega}{2\pi},\frac{g}{2} - i \frac{\beta
\omega}{2\pi} )\, ,
\end{eqnarray}
where $B(x,y)$ is the Euler Beta function. In the zero temperature
limit, formula (\ref{spectral}) reduces to the standard
zero-temperature chiral Luttinger liquid spectral function~:
 \begin{equation}
A_{R,L}(k,\omega)= \alpha^{g-1}\frac{\displaystyle
2\pi}{\displaystyle \Gamma(g)}  |k \mp k_{f}|^{g-1}  \delta
(\omega + v( k_{f} \mp k)).
\end{equation}

\begin{figure}[!htbp]
\begin{center}\includegraphics[width=3.25in,
  keepaspectratio]{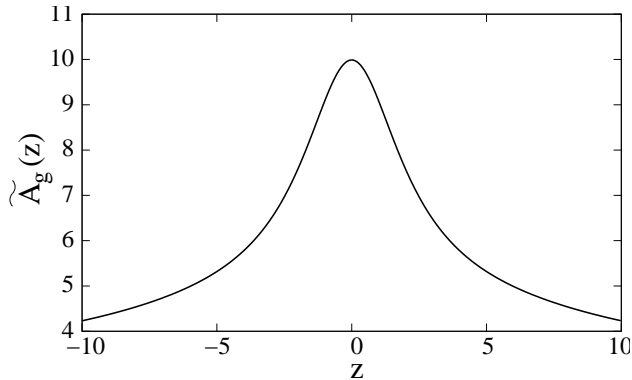}
\end{center}
\caption{The scaling function $\widetilde{A}_g$(z)of a
chiral Luttinger liquid as a function of the scaling variable
$z={\displaystyle \omega}/{\displaystyle T}$ in the case
g=0.7. } \label{spectralfn}
\end{figure}

The spectral function can reexpressed to show its scaling properties
(we define the Fermi energy $\varepsilon_f =vk_f$)~:
\begin{equation}\label{scale}
A_{R,L}(k,\omega)= 2(2\pi)^{g-1}\left(\frac{T}{\varepsilon_f}\right)^{g-1}
\widetilde{A}_g(\omega/T)\,\,
\delta (\omega +\varepsilon_f \mp vk),
\end{equation}
with the scaling function~:
\begin{equation}\label{auxF}
\widetilde{A}_g(z)=(\alpha k_f)^{g-1}\cosh(z/2)\,\, B( \frac{g}{2}
+ i \frac{z}{2\pi},\frac{g}{2} - i \frac{z}{2\pi} ).
\end{equation}
We note that the $\omega/T$ scaling is destroyed when $g\neq 1$
by the overall factor $\left({T}/{\varepsilon_f}\right)^{g-1}$.
The scaling function $\widetilde{A}_g$ is plotted in Fig.~\ref{spectralfn}.

\section{Results}
Our formula Eq.(\ref{IV}) yields the value of the tunneling
current for any barrier length $L$, interaction parameter $g$,
finite applied voltage and temperature. It is convenient to define a
dimensionless  voltage
$\overline{v}={\displaystyle eV}/({\displaystyle 2\varepsilon_{f}})$.
We will concentrate on the zero temperature
limit. The tunnel current can be written as~:
\begin{equation}\label{Zero}
I=I_0\,\, (k_f L)^2
\int_{-\overline{v}}^{+\overline{v}}\, du \, \frac{\displaystyle
\sin^{2}[(1+u)k_fL]}{\displaystyle
(1+u)^{2}(k_fL)^{2}}(\overline{v}^2-u^2)^{g-1},
\end{equation}
where we have defined an overall current scale
$I_{0}={\displaystyle e\mathcal{T}^{2}(\alpha k_{f})^{2g-2}}/{\displaystyle
(8\pi\varepsilon_{f}\Gamma(g)^{2})}$.
It is not possible to compare directly currents with and without interactions
because the current scale $I_0$ is a function of $g$. More precisely
the microscopic cut-off appears through the dimensionless combination
$\alpha k_f$ when $g\neq 1$.
In the $g=1$ case, one can perform the remaining integral in Eq.(\ref{Zero})
in terms of the Sine integral~\cite{Abramowitz}~:
\begin{eqnarray}\label{freecase}
\frac{I}{I_0}= &&k_f L \,\,
[Si(2k_f L(\overline{v}+1))+Si(2k_f L(\overline{v}-1))]\\
\nonumber
&&-\frac{1-\cos(2k_f L(\overline{v}+1))}{2(\overline{v}+1)}
-\frac{1-\cos(2k_f L(\overline{v}-1))}{2(\overline{v}-1)}.
\end{eqnarray}

\subsection{Non-resonant case}

We first discuss the case when $k_f\neq 0$.
This means that, in the absence of bias, either the two Fermi
points are both below the maximum tunneling point $k=0$
where the dispersion relations do cross
or both are above the $k=0$. So tunneling is suppressed till a finite bias.
This can be seen in formula Eq.(\ref{Zero}).
If we set the barrier
length to infinity then the sine function in the integrand of the
tunnel current Eq.(\ref{Zero}) peaks to a delta function. The
current is then nonzero only beyond a bias threshold given by
$|eV|>2vk_f$. This is simply due to the conservation of momentum
in the limit $L\rightarrow\infty$ which severely restricts
tunneling. This threshold is smoothed out for finite $L$ values
and disappears in the point contact limit $L\rightarrow 0$.
We will refer to the generic case $k_f\neq 0$ as being "non-resonant".
On the contrary, for $k_f =0$ tunneling is allowed for infinitesimal bias,
the "resonant" case treated in the next section.
In a realistic set-up one needs a special fine-tuning of the magnetic field
to reach the resonant case. In the experiment of Kang et al, there is an extended
region of $k_f$, i.e. of magnetic field where there is sizeable tunneling
for infinitesimal bias as revealed by a zero-bias peak in the conductance.
Such an extended range of tunneling is out of reach of the present model
where there is a single point satisfying the resonance condition.

In
fact, the tunnel current is O($L^0$) below the threshold and grows
as O($L$) above. Below the threshold, the $L\rightarrow\infty$
limit leads to~:
\begin{equation}\label{below}
\frac{I}{I_0} =\left(\frac{eV}{2\varepsilon_f}\right)^{2g-1}
\frac{\sqrt{\pi}\Gamma(g)}{2\Gamma(g+1/2)}\\\nonumber\times
{}_{2}F_{1}(1,\frac{3}{2},g+\frac{1}{2};\left(\frac{eV}{2\varepsilon_f}\right)^2),
\end{equation}
where ${}_{2}F_{1} $ is the hypergeometric function~\cite{Abramowitz}.
Above threshold, we find the simple formula~:
\begin{equation}\label{above}
\frac{I}{I_0}=\pi\,\, k_fL\,\, \left(\left(\frac{eV}{2\varepsilon_f}\right)^2-1\right)^{g-1}.
\end{equation}
\begin{figure}[!htbp]
\begin{center}\includegraphics[width=3.25in,
  keepaspectratio]{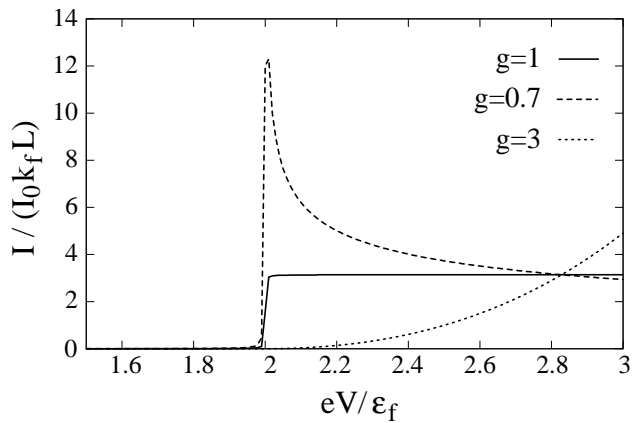}
\end{center}
\caption{Tunneling current  vs
${\displaystyle eV}/{\displaystyle \varepsilon_{f}}$ for three
values of $g$~: $g=3$, (dotted line) $g=1$ (solid line), $g=0.7$
(dashed line) for a barrier of length $k_fL=1000$. Since the normalization $I_0$
of the currents is a function of $g$, no comparison can made between
the absolute value of these $I-V$ curves.}
\label{infiniteL}
\end{figure}

Some $I-V$ curves are displayed in fig.~(\ref{infiniteL}) for a
very long barrier, $k_f L\gg 1$. In the FQHE case, we have taken
the Luttinger parameter $g=3$ for tunneling between two $\nu =1/3$
liquids. The case $g=1$ refers to free fermions, here there is a
saturation of the tunnel current, i. e. $I$ becomes O($L$)
independent of $V$. Finally, we have drawn the $I-V$ curve for a
Luttinger parameter less than one, possibly relevant to the case
of IQHE edge modes with interedge interactions. There is a sharp
current spike at threshold. The current indeed behaves as $\sim
(eV-2\varepsilon_f)^{g-1}$ close to threshold in the large-$L$
limit. For $V\rightarrow 0$, the current behaves as $V^{2g-1}$.
\begin{figure}[!htbp]
\includegraphics[width=3.in, keepaspectratio]{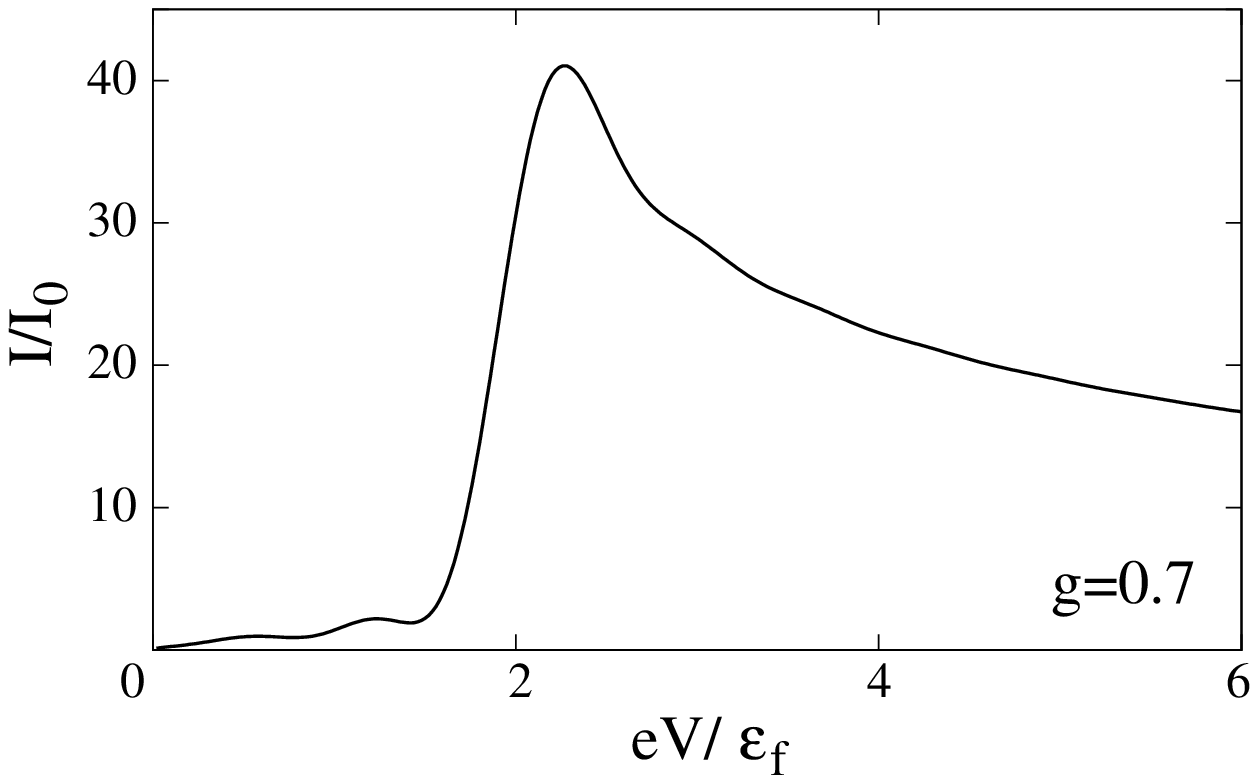}
\vspace{0pt}
\includegraphics[width=3.in, keepaspectratio]{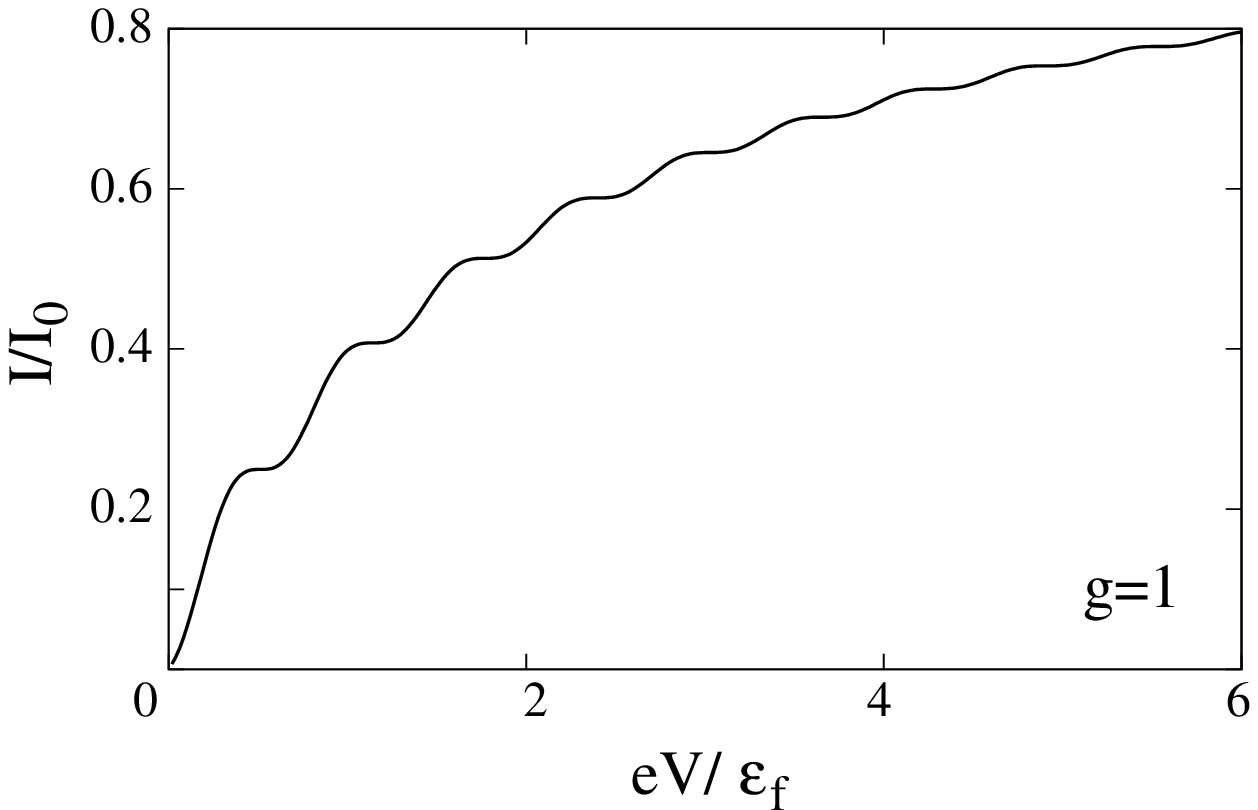}
\vspace{0pt}
\includegraphics[width=3.in, keepaspectratio]{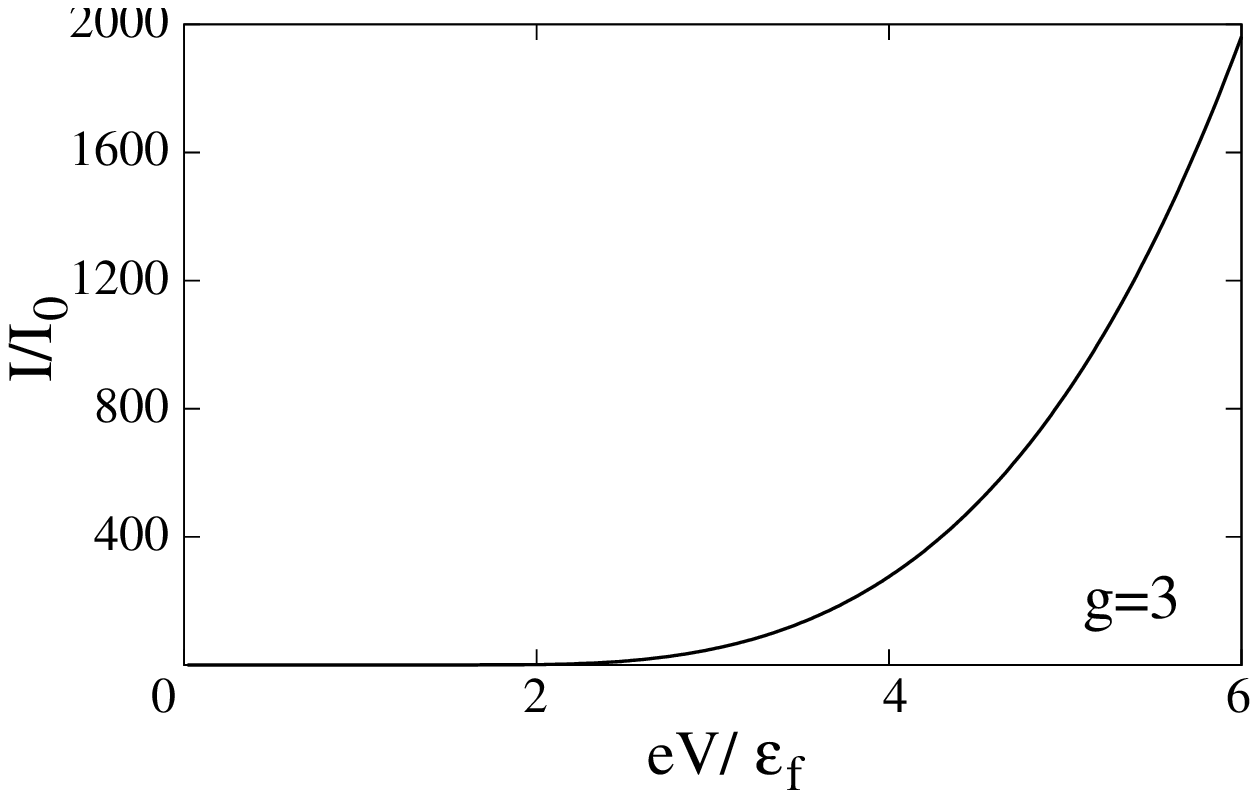}
\caption{Tunneling current  vs
${\displaystyle eV}/{\displaystyle \varepsilon_{f}}$ for three
values of $g$~: $g=0.7$, $g=1$ and $g=3$ from top to bottom
 for a barrier of length $k_fL=10$.}
\label{IV10}
\end{figure}

\begin{figure}[!htbp]
\begin{center}\includegraphics[width=3.25in,
  keepaspectratio]{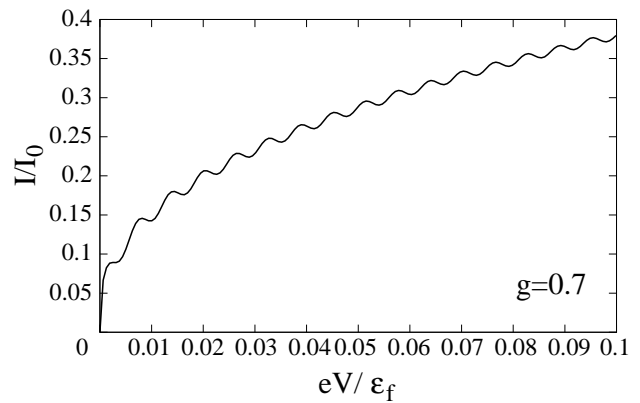}
\end{center}
\caption{Tunneling current vs ${\displaystyle eV}/{\displaystyle
\varepsilon_{f}}$ for the interacting parameter $g=0.7$ for a
finite barrier length $k_f L=1000$. } \label{IV1}
\end{figure}

For a finite length barrier with $k_fL$ of order unity, the kinematic
singularities are smoothed out. Typical curves are presented in
Fig.(\ref{IV10}).
The behavior for $V$ small is
unchanged but there are additional oscillations due to the
diffraction pattern generated by the barrier acting like a slit.
We note that in the case with $g$ less than one, the singular behavior
for small voltage $\propto V^{2g-1}$ shows up only for small voltage.
In fig.(\ref{IV1}) is presented a close-up of the case $g=0.7$ at very
small bias~:
here one can observe the nonlinearity typical of a Luttinger
liquid and the additional interference pattern. The period of the
oscillations is $2\pi /(k_fL)$ in terms of the dimensionless voltage
$eV/(2\varepsilon_f)$.

\begin{figure}[!htbp]
\begin{center}\includegraphics[width=3.25in,
  keepaspectratio]{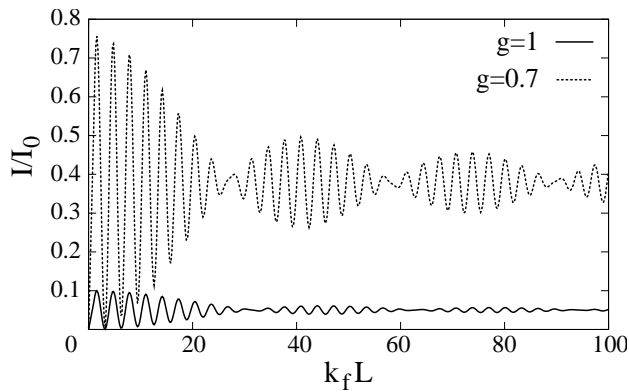}
\end{center}
\caption{Tunneling current in unit  of $I_{0}$ vs length of the
barrier for the interacting parameter $g=1$ (solid line) and
$g=0.7$ (dotted line). The bias is fixed at $eV/(2\varepsilon_f)=0.1$.}
\label{IVL}
\end{figure}

Fig.~\ref{IVL}  shows the tunneling current vs the length of the
barrier for a  fixed voltage $eV / (2\varepsilon_{f})=0.1$ and several
values of the interaction parameter $g$. There is a saturation for
$L$ larger than hundreds of the Fermi wavelength $\lambda_{f}$, as
discussed above, because the voltage bias is below the threshold.
 There is also a beating pattern in both cases.
The fast oscillations are due to the interference of L versus the
Fermi wavelength while the slow oscillations are ruled by the
voltage scale introduced in the problem by the bias voltage
$eV/(2\varepsilon_f) $ set to 0.1.
We have omitted the case $g=3$ because it has a similar shape but with a very small
current in units of $I_0$.

\subsection{Resonant case}

\begin{figure}[!htbp]
\begin{center}\includegraphics[width=3.25in,
  keepaspectratio]{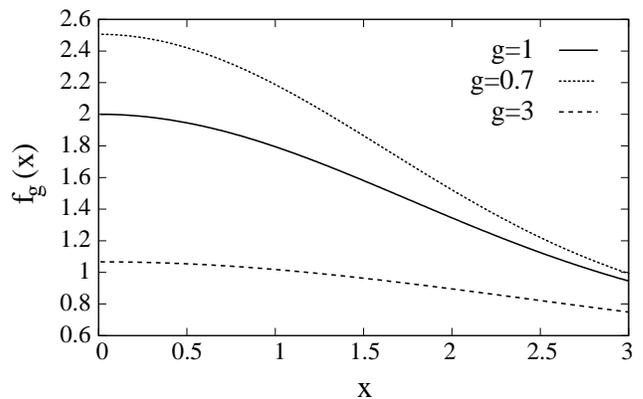}
\end{center}
\caption{The auxiliary function governing the crossover between the regimes of
long and short barriers for values of the parameter $g$.} \label{fg}
\end{figure}

\begin{figure}[!htbp]
\includegraphics[width=2.5in, keepaspectratio]{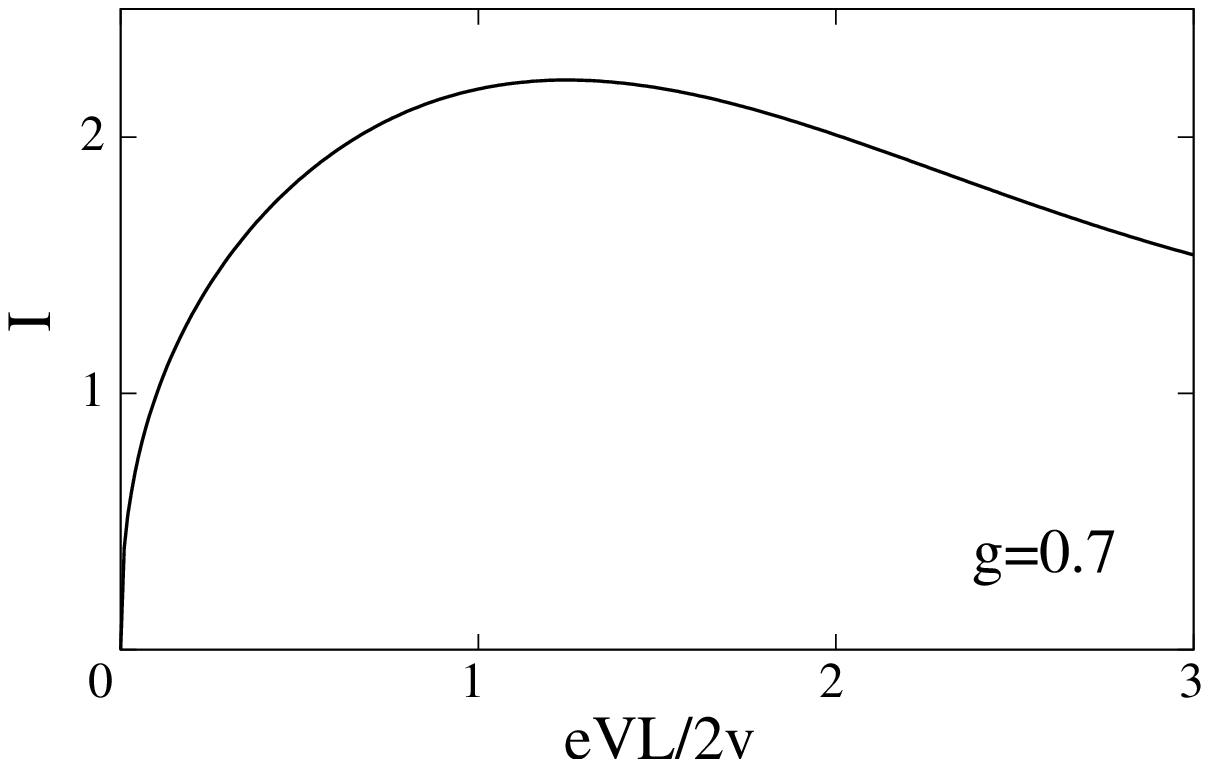}
\vspace{0pt}
\includegraphics[width=2.5in, keepaspectratio]{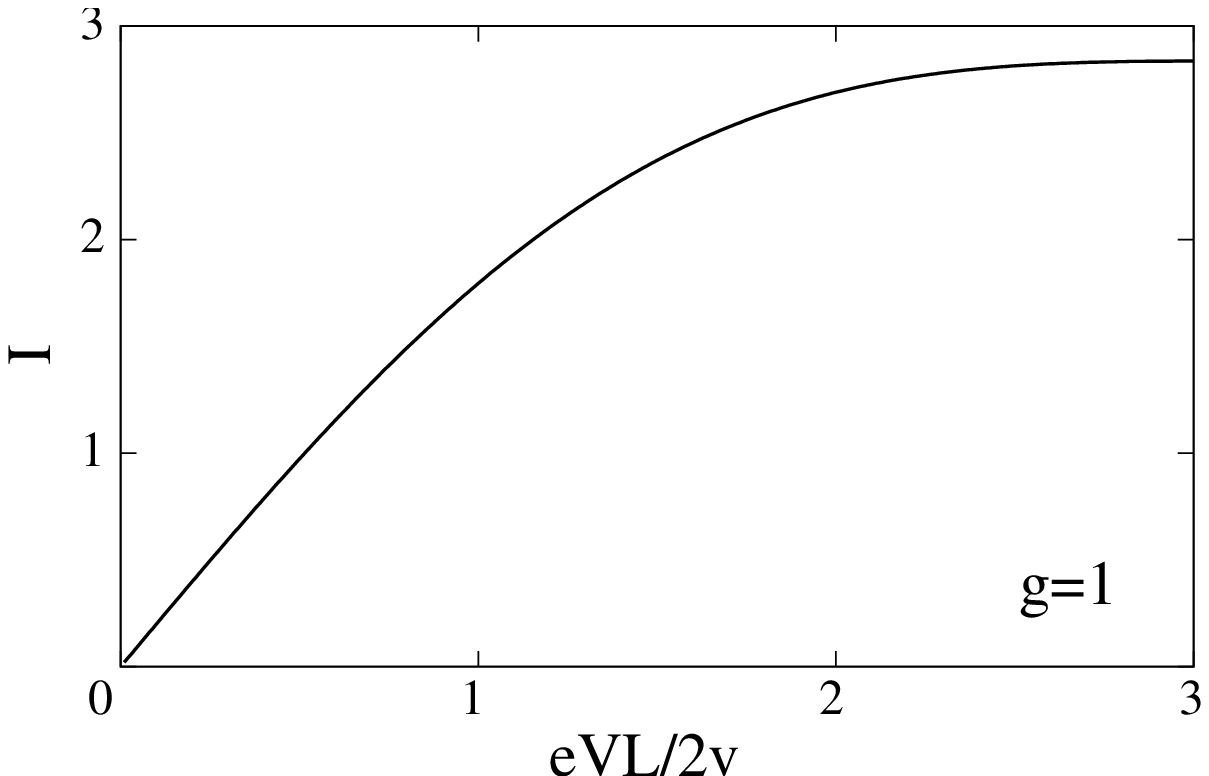}
\vspace{0pt}
\includegraphics[width=2.5in, keepaspectratio]{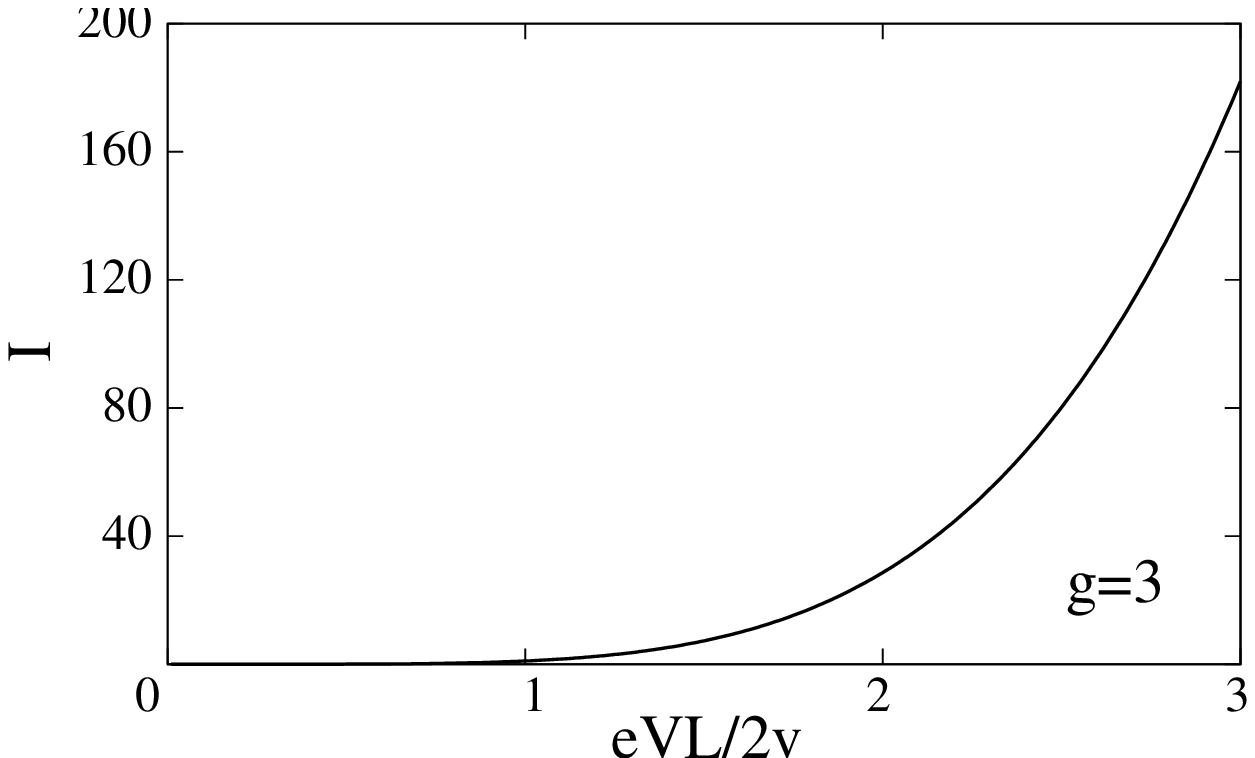}
\caption{Tunneling current  vs
bias $eVL/(2v)$ at resonance for three
values of $g$~: $g=0.7$, $g=1$ and $g=3$ from top to bottom.}
\label{ResoIV}
\end{figure}

We now turn to the case $k_f =0$ which can be obtained by tuning
the external applied magnetic field.
This means that in the absence of bias the left and right Fermi levels
exactly coincide at $k=0$.
The tunnel current from our
perturbative calculation is now simpler because the Fermi energy disappears
from the problem. As a consequence we find~:
\begin{equation}\label{Reso}
I=\frac{e\mathcal{T}^2}{\Gamma (g)^2}\frac{L^2}{8\pi\alpha v}
\left(\frac{\alpha eV}{2v}\right)^{2g-1}\,\, f_g(\frac{eVL}{2v}),
\end{equation}
where we define the auxiliary function $f_g(x)$ by~:
\begin{equation}\label{auxf}
f_g (x)=\int^{+1}_{-1}dt\,\, (1-t^2)^{g-1}
\frac{\sin^2(xt)}{(xt)^2}.
\end{equation}
This function decreases as $\pi/x$ for large $x$ and any value of
$g$ and has a nonzero value at the origin. This means that
 $I\propto V^{2g-1}$ if $eVL/v\ll 1$ while
$I\propto V^{2g-2}$ if $eVL/v\gg 1$. For a typical
$g=0.7$ value this means that the current should first rise as
$V^{0.4}$ before going down at larger voltage as $V^{-0.6}$. The
function that governs the crossover is shown in Fig.(\ref{fg}).
Some $I-V$ curves are displayed in Fig.(\ref{ResoIV}).
While the Luttinger liquid nonlinearities are still present, the
oscillations formerly due to the presence of the scale $k_f$ no
longer exist. The $g=0.7$ curve shows that strong interedge interactions
in a situation of weak tunneling lead to a diverging conductance at zero bias.

Finally we note for completeness that our study may be extended at nonzero temperature
by using the finite-temperature spectral functions. Using a dimensionless temperature
$z=\beta\varepsilon_f$, we have~:
\begin{eqnarray}
\label{IVft} &I&=\frac{\displaystyle e
\mathcal{T}^{2}}{\displaystyle 8\pi\varepsilon_{f}} \frac{
(\alpha k_{f})^{2g-2}}{\Gamma(g)^{2}}(\frac{\displaystyle 2\pi
}{\displaystyle z})^{2g-2}\frac{\displaystyle \Gamma(g)^{2}
}{\displaystyle 4\pi^{2}} (k_fL)^2 \times
\\\nonumber
&&\int \, du \,
\frac{\displaystyle \sin^{2}[(1+u)k_fL]}{\displaystyle
(1+u)^{2}(k_fL)^{2}}e^{zu}(1+e^{-z(u+\overline{v})})(1+e^{-z(u-\overline{v})})\\
\nonumber &&\times B\left(\frac{g}{2} + i
\frac{z}{2\pi}\left(u+\overline{v}\right),\frac{g}{2}-i
\frac{z}{2\pi}\left(u+\overline{v}\right)\right)
\\\nonumber
&&\times B\left(\frac{g}{2}
+ i \frac{z}{2\pi}\left(\overline{v}-u\right),\frac{g}{2}-i
\frac{z}{2\pi}\left(\overline{v}-u\right)\right).
\end{eqnarray}
We expect thermal rounding of the oscillatory features as well as the threshold behavior
in the non-resonant case.

\section{Conclusions}
In this paper we have studied the tunneling between two
counterpropagating edges modes pertaining to different electron
gases by means of a perturbative expansion. We find the
characteristic nonlinearities of the Luttinger liquid properties.
There are additional oscillations due to the finite extent of the
barrier. Our study applies to the case of tunneling between two FQHE
fluids for $\nu =1/m$ and also applies to the case of tunneling
between edges when there is strong interactions throughout the
barrier (but weak tunneling) in which case the characteristic
exponent entering the spectral functions is no longer quantized by
the bulk physics. It remains to be seen if samples with the
correct properties can be artificially produced. We note that the
present experiment by Kang et al is apparently in the regime of
strong tunneling (even if the conductance throughout the barrier
is far below the expected Landauer value).


\begin{acknowledgments}
We thank  S. M. Girvin for useful correspondence.
We also thank the members of the mesoscopic physics group at ENS
for fruitful discussions.
\end{acknowledgments}


\end{document}